\documentclass{article}

     \PassOptionsToPackage{numbers, compress}{natbib}
 \usepackage[final]{tackling_climate_workshop_style}
 
\usepackage{layouts}

\usepackage[utf8]{inputenc} % allow utf-8 input
\usepackage[T1]{fontenc}    % use 8-bit T1 fonts
\usepackage{hyperref}       % hyperlinks
\usepackage{url}            % simple URL typesetting
\usepackage{booktabs}       % professional-quality tables
\usepackage{amsfonts}       % blackboard math symbols
\usepackage{nicefrac}       % compact symbols for 1/2, etc.
\usepackage{microtype}      % microtypography
\usepackage{xcolor}
\usepackage{graphicx}
\usepackage{bbm}
\usepackage{amsmath}
\usepackage{booktabs}

\title{Improving the predictions of ML-corrected climate models with novelty detection}

\author{%
  Clayton Sanford\\
  Department of Computer Science\\
  Columbia University\\
  New York, NY 10027 \\
  \texttt{clayton@cs.columbia.edu} \\
  \And 
  Anna Kwa \\
  Climate Modeling \\
  Allen Institute for AI \\
  Seattle, WA 98103 \\
  \texttt{annak@allenai.org}
  \And 
  Oliver Watt-Meyer \\
  Climate Modeling \\
  Allen Institute for AI \\
  Seattle, WA 98103 \\
  \texttt{oliverwm@allenai.org}
  \And 
  Spencer Clark \\
  Climate Modeling \\
  Allen Institute for AI \\
  Seattle, WA 98103 \\
  \texttt{spencerc@allenai.org}
  \And 
  Noah Brenowitz \\
  Climate Modeling \\
  NVIDIA \\
  Santa Clara, CA 95051 \\
  \texttt{nbrenowitz@nvidia.com }
  \And 
  Jeremy McGibbon \\
  Climate Modeling \\
  Allen Institute for AI \\
  Seattle, WA 98103 \\
  \texttt{jeremym@allenai.org}
  \And 
  Christopher Bretherton \\
  Climate Modeling \\
  Allen Institute for AI \\
  Seattle, WA 98103 \\
  \texttt{spencerc@allenai.org}
}

\begin{document}

\maketitle

\begin{abstract}
While previous works have shown that machine learning (ML) can improve the prediction accuracy of coarse-grid climate models, these ML-augmented methods are more vulnerable to irregular inputs than the traditional physics-based models they rely on.
Because ML-predicted corrections feed back into the climate model’s base physics, the ML-corrected model regularly produces out of sample data, which can cause model instability and frequent crashes.
This work shows that adding semi-supervised novelty detection to identify out-of-sample data and disable the ML-correction accordingly stabilizes simulations and sharply improves the quality of predictions.
We design an augmented climate model with a one-class support vector machine (OCSVM) novelty detector that provides better temperature and precipitation forecasts in a year-long simulation than either a baseline (no-ML) or a standard ML-corrected run.
By improving the accuracy of coarse-grid climate models, this work helps make accurate climate models accessible to researchers without massive computational resources.
% Goal is to have accurate, robust, resource-efficient climate models. 
% Common strategy for doing so is an ML-nudged climate model, which succeeds in reducing biases in coarsened climate models.
% Problem: ML is unreliable and may create unphysical pushes that lead to instability. 
% Is there any way to reap the accuracy and resource-efficiency of ML-nudged coarse climate models without suffering from instability? 
% Yes; models perform poorly when out-of-sample, and novelty detection algorithms allow the model to use ML approach when it's likely to help and ignore it when not.
% We improve most metrics of model performance coarse model performance by introducing wind tendencies to the model, whose inclusion otherwise introduces instability.
\end{abstract}

\section{Introduction}

Accurate climate models are essential for diagnosing the general trends of climate change and predicting its localized impacts.
Given finite resources, having \textit{computationally efficient} models is also important to assess climate policies by making simulations cheap and easy. 
Previous works \cite{Krasnopolsky2010-nn,bb19, wbc21, yo20, rpg18} have suggested that augmenting physics-based climate models with machine learning can reduce bias and improve the overall skill of coarse climate models, while sometimes introducing instability.
This work draws on the idea of using a compound parameterization  \cite{Krasnopolsky2008-bw, Song2021-ug} to mask ML models with high uncertainty and builds on those ML-corrected models by incorporating \textit{out-of-sample detection}.
Our approach adds stability and outperforms these past approaches (specifically, \cite{wbc21, bhk22}) on temperature and precipitation metrics.

We model the atmosphere as a system of partial differential equations (PDEs).
The atmospheric state is modeled as $X = (x_1, \dots, x_N) \in \mathbb{R}^{d \times N}$, a three-dimensional grid of $N$ latitude/longitude coordinates with $d$-dimensional column vectors of air temperature, specific humidity, and other fields.
% We let 
% $d$-dimensional vectors representing various climate measurements over a particular latitude/longitude coordinate (e.g. the air-temperature or specific humidity at a particular level of pressure) at $N$ grid points spaced out around the globe, which can be collected as $X = (x_1, \dots, x_N) \in \mathbb{R}^{d \times N}$.
The state of a particular column $x_i \in \mathbb{R}^d$ evolves over time as
\begin{equation}\label{eq:base}
\frac{dx_i}{dt} = f_i(X, t)
\end{equation}
for some fixed $f_i$ derived from physically-based assumptions; we refer to this as the \textit{baseline model}.
The size of $N$ corresponds to the grid resolution; large $N$ yields more accurate but computationally expensive simulations.

While accuracy penalties due to a loss of resolution are expected for small $N$, coarse-grid simulations are additionally biased by poor representations of subgrid-scale processes like thunderstorms and cloud radiative effects \cite{zw06, wybp18}.
ML is an appealing way to de-bias this coarse climate model by predicting and compensating for its error.
Put precisely, the \textit{ML-corrected model} is
\begin{equation}\label{eq:prog}
    \frac{dx_i}{dt} = f_i(X, t) + g(x_i, q_i; \theta),
\end{equation}
where $g(\cdot; \theta): \mathbb{R}^{d+3} \to \mathbb{R}^d$ is a learned function with parameters $\theta$ that predicts corrective tendencies from the column, $x_i \in \mathbb{R}^d$, and its insolation, surface elevation, and latitude $q_i \in \mathbb{R}^3$.
The ML correction enables the baseline to better approximate a reference fine-grid model while maintaining the underlying physics as the core of the modeling approach \cite{bb19, wbc21}.

While ML-based models frequently improve overall error, these models---especially deep neural networks---are often not robust, meaning they perform poorly with \textit{out-of-sample} data that lies outside the training distribution.
In online application (where predictions are fed back into the model repeatedly for a simulation) of these models, ML model errors accumulate in time and overwhelm the damping mechanisms of the baseline physics \cite{bhc20}.
In past works \cite{bhk22}, letting $x_i$ represent a vertical column of air temperature and specific humidity values resulted in an accurate and stable model, but including horizontal winds in $x_i$ caused the model to crash or be more inaccurate for certain fields within a few simulated weeks.
Other works \cite{cbh22} have stabilized ML-corrected climate models by tapering upper-atmospheric outputs to zero and removing upper-atmospheric inputs when learning $g$, but this approach has not been applied to models with wind tendencies.

This poses a dilemma: 
By omitting the wind tendencies from $g$, the model is unable to incorporate relevant climate information into its predictions.
Yet including the wind tendencies introduces new instabilities.
We fix this by employing semi-supervised \textit{novelty detection} to predict when a column $x_i$ belongs to the training distribution of $g$ and suppress the tendencies of the ML model if not.
% Can we obtain the best of all worlds: an efficient, stable model that corrects biases to the horizontal winds and achieves a more skillful prediction?
% This work answers the question in the affirmative by employing semi-supervised \textit{novelty detection} to predict when a column $x_i$ belongs to the training distribution of $g$ and suppress the tendencies of the ML model if not.
Our model has the form 
\begin{equation}\label{eq:nd}
    \frac{dx_i}{dt} = f_i(X, t) + \eta(x_i; \rho)g(x_i, q_i; \theta),
\end{equation}
for a novelty detector $\eta(\cdot; \rho): \mathbb{R}^d \to [0, 1]$.
A properly tuned $\eta$ improves coarse climate model temperature and precipitation forecasts for at least a year.

% We employ out-of-sample detection algorithms to predict when an input does not belong to the training distribution and suppress the ML-based nudge if so.
% By exploring different approaches, we can tweak the parameterizations to interpolate between the prognostic ML-aided run (everything in sample) and the baseline no-ML run (nothing in sample).
% The proper choice of OOS algorithm allows us to reap the best of both worlds: improvements on the metrics of the baseline and prognostic runs without the previous stability issues.

% \subsection{Related work}

% \red{Discuss previous papers in and out of group \cite{wbc21, bhk22, bb19}.
% Discuss prevalence of ML nudging and cite concern about out-of-sample issues in literature.}

\section{Methodology}
\subsection{ML-corrected climate models and data}
We consider two neural networks for modeling the ML-corrected tendencies.
$g_{\mathrm{Tq}}$ corrects vertical columns with 79 pressure levels containing only air temperatures (T) and specific humidities (q); thus, $d = 2 \cdot 79$.
$g_{\mathrm{Tquv}}$ additionally corrects eastward and northward wind velocities (u, v); $d = 4 \cdot 79$.
We train the these corrective functions to predict an observed ``nudging'' vector between a pre-existing fine-grid simulation (with much larger $N$) and a simulated coarse-grid run \cite{wbc21, bhk22}.
We use a dataset \smash{$\{x^{(t)}_i \in \mathbb{R}^d: t \in [T], i \in [N]\}$} with $T$ time steps generated by the same simulation as the training set to train the novelty detector $\eta(\cdot, \rho)$.
These models are described in greater depth in Appendix~\ref{asec:training}.

% \red{more technical details of these models will go in the Appendix~\ref{asec:prog-param}

% Introduce problem as dynamical systems with learned terms.
% Temperature, humidity, horizontal winds as variables.
% Learn a corrective nudge for each as a supervised ML problem, with corrections from baseline coarse run to baseline fine run as ground truth.

% Discuss SOTA of ML-corrected climate models.
% What metrics do we care about and why?
% Discuss difficulty in improving metrics like total water path.

% Challenge: this is an online approach and minor mistakes can cascade into larger ones.
% Physics-based models are stable and good at self-correction, but ML is not, especially when outside of training set.
% This is particularly bad for neural network-based predictors.
% }

\subsection{Novelty detection}
% We 
%moderate the influence of the ML corrections by assuming
% assume that the ML model should behave reasonably if a sample belongs to the training distribution.
The novelty detector $\eta$ predicts whether a column $x$ belongs within the support of the training set.
If so, then we let $\eta(x; \rho) = 1$ to take full advantage of the learned correction $g(x, q; \theta)$; otherwise, we ignore $g(\cdot; \theta)$ by setting $\eta(x; \rho) = 0$.\footnote{The body of the paper only considers novelty detectors with sharp thresholds (i.e. $\eta(\cdot; \rho): \mathbb{R}^d \to \{0, 1\}$). See Appendix~\ref{asec:threshold} for an examination of continuous-valued novelty detectors.}

\textit{Novelty detection} is a well-studied semi-supervised learning problem about estimating the support of a dataset using only positive examples \cite{ha04}. 
We frame the problem as novelty detection rather than \textit{outlier detection} (an unsupervised problem with mixture of in-distribution and out-of-distribution samples) or standard two-class supervised classification because we have no dataset of representative out-of-distribution samples.
% While prognostic runs routinely return samples that do not belong to the training set, future out-of-sample are not guaranteed to resemble those.
% A novelty detection framework establishes no distributional model of what out-of-distribution samples might look like; it requires only a model of the support of the distribution.
There are many known approaches to novelty detection, including local-outlier factor \cite{bkns00}, $k$-means clustering
\cite{ntc99}, and minimum-volume ellipsoid estimation \cite{vr09}.
Our exploratory work considers two of these approaches: a simple ``min-max'' novelty detector and a one-class support vector machine (OCSVM).
For each of these we consider novelty detectors $\eta_{\mathrm{T}}$ with 79-dimensional temperature vectors as input and $\eta_{\mathrm{Tq}}$ with 158-dimensional combined temperature and specific humidity vectors.

\paragraph{Naive ``min-max'' novelty detector}
The min-max novelty detector considers the smallest axis-aligned hyper-rectangle that contains all training samples and categorizes any sample outside the rectangle as a novelty.
Put concretely, 
\[\eta_{\mathrm{minmax}}(x; (x_{\min}, x_{\max})) = 
\begin{cases}
    1 & \text{if } x_{k} \in [x_{\min,k}, x_{\max,k}] \  \forall k \in [d], \\
    0 & \text{otherwise,}
\end{cases}\]
for $x_{\min, k} = \min_{i, t} x^{(t)}_{i, k}$ and $x_{\max, k} = \max_{i, t} x^{(t)}_{i, k}$ as the minimum and maximum over the training data of the $k$th feature.
% This requires only $O(NTd)$ and $O(d)$ time complexity respectively to train and predict. 
While efficient, this novelty detector is unable to identify irregularities within the bounding box.

% Why might this not work?
% \begin{itemize}
%     \item Temperature profiles vary a lot, and models frequently mess up parts of columns.
%     See issues with upper- and lower-atmosphere diverging
%     \item This method puts outsize importance on the poles and tropics: since shared representation across the world, all of the extremal points for temperatures will be in those places. A more sophisticated representation of the distribution may have other boundaries.
% \end{itemize}

\paragraph{One-class support vector machine (OCSVM)}

The one-class SVM algorithm of \cite{spss01} repurposes the SVM classification algorithm to estimate the support of a distribution by finding the maximum-margin hyperplane separating training samples from the origin. 
The OCSVM has been applied to novelty detection for genomics \cite{shsg17}, video footage \cite{avja18}, propulsion systems \cite{tnthz19}, and the internet of things \cite{ykf21}.
We normalize each input $x_i$ and lift it to the infinite-dimensional feature space $\phi(x_i)$ corresponding to the radial basis function (RBF) kernel $\kappa_\gamma(x, x') = \exp(-\gamma\|x - x'\|_2^2)$.
We use the novelty detector 
\[\eta_{\mathrm{OCSVM}}(x; (\alpha, \nu, \gamma)) = 
\begin{cases}
    1 & \text{if } \sum_{t=1}^T \sum_{i=1}^N \alpha_{t,i} \kappa_\gamma(x, x_{i}^{(t)})\geq \nu , \\
    0 & \text{otherwise,}
\end{cases}\]
whose weights $\alpha_i \geq 0$ are determined by a quadratic program based on the training data and whose sensitivity is set by cutoff $\nu > 0$. 
The prediction rule depends exclusively on the number of \textit{support vectors}, or training samples \smash{$x_{i}^{(t)}$} with $\alpha_{t,i} > 0$.
To obtain a robust and computationally efficient novelty detector, we restrict the model at have most a $10^{-4}$ fraction of samples as support vectors.
% We use the \textit{radial basis function (RBF)} kernel $\kappa(x,x') = \exp(-\gamma \|x - x'\|^2)$.
Smaller values of $\gamma$ correspond to novelty detectors that with highly smoothed support estimations that may be larger than necessary, while large $\gamma$ provides a smaller and higher variance region.
Our experiments train and evaluate numerous novelty detectors for several choices of $\gamma$ and $\nu$.

% \red{
% write down formulation/equations 
% }

% Anything on the boundary is a support vector.
% $\eta$ parameter is a regularization term that restricts how many support vectors can be chosen. (Here, we choose $\eta$ to get roughly \red{number here} SVs, for computational reasons.)
% Efficiency depends on number of support vectors, since resulting classifier relies on kernel computed with those SVs.

% Q1: How well does any approach work at modulating issues?
% Q2: Is there a benefit to nonlinear approaches like OCSVM with RBF?

\section{Results}\label{sec:results}

% \subsection{Comparison of runs with and without novelty detection}
Because there is no ground-truth verdict about whether a data point is out-of-distribution, we evaluate our novelty detectors by incorporating $\eta(\cdot; \rho)$ into the coarse grid model, numerically simulating equation~\eqref{eq:nd} for one year at 15-minute time increments at C48 resolution, 
%with the FV3GFS model \cite{zlcmcr19}, 
and comparing the predicted atmospheric states \smash{$\hat{x}_{i}^{(t)}$} to \smash{$x_{\mathrm{fine}, i}^{(t)}$} using the root mean square error (RMSE) of three time-averaged diagnostics (see equation~\eqref{eq:rmse}): near-surface air temperatures at 850hPa of pressure (\textbf{T}), surface precipitation rate (\textbf{SP})\footnote{Current climate models have less confident predictions of regional shifts in precipitation than of surface temperatures; contrast sections B.2.1 from B.3.1 of \cite{icpc21}.}, and precipitable water (\textbf{PWAT})\footnote{PWAT is the total mass of water contained in a vertical atmospheric column per cross-sectional area and is closely related to the regional precipitation rate \cite{bpb04}.}.
% \begin{itemize}
%     \item Root mean square error (RMSE) of time-averaged near-surface air temperatures at 850hPa of pressure (\textbf{t850}):
%     % If $x_{i, \mathrm{temp-850}}^{\mathrm{model}(t)}$ and $x_{i, \mathrm{temp-850}}^{\mathrm{true}(t)}$ are the estimated and observed temperatures at 850kPa of pressure, and , we compute 
%     \[\sqrt{ \sum_{i=1}^N a_i \left(\frac1T \sum_{t=1}^T \left(\hat{x}_{ i, \mathrm{t(850hPa)}}^{(t)} - x_{\mathrm{fine},i, \mathrm{t(850hPa)}}^{(t)}\right)\right)^2},\]
%     where $a_i \in \mathbb{R_+}$ is the fraction of Earth's surface area represented by cell $i$.
%     \item RMSE of the total surface precipitation (\textbf{SP}). \red{more uncertainty about precipitation wrt climate change; focus on local tropospheric metrics}
%     % This can be estimated as a function of the the columns $x_{i, \mathrm{temp-850}}^{\mathrm{model}(t)}$ and measured against observed precipitation levels.
%     \item RMSE of the precipitable water (\textbf{PWAT}), the total volume of water contained in a vertical atmospheric column. %\red{add justification and citation, maybe nudged-to-PIRE?}
%     % \item RMSE over five days of the specific humidity at 500kPa. %\red{why this one?}
% \end{itemize}
Table~\ref{tab:metrics} compares the performances of seven global simulations.
The first is the baseline simulation of equation~\eqref{eq:base}; the next two are ML-corrected runs from equation~\eqref{eq:prog} without and with wind tendency corrections; and the remaining four simulations include novelty detection from equation~\eqref{eq:nd} and differ in the choice of novelty detector and the inputs to $g$ and $\eta$.
The OCSVMs use RBF parameter $\gamma = \frac4{79}$ and cutoff 
\begin{equation}\label{eq:nu}
    \nu = \min_{i', t'} \sum_{t} \sum_i \alpha_{t,i} \kappa_\gamma(x_{i'}^{(t')}, x_{i}^{(t)})
\end{equation} set to be the smallest OCSVM score observed in the training set.

\begin{table}[]
    \centering
    \caption{The RMSE scores of time-averaged metrics and novelty detection rates for year-long runs. The third run crashed after 38 days (denoted with a star), so its metrics are averaged over 38-days rather than one year. The ``\% Novelty'' column represents the fraction of columns of the simulated run without ML-corrections. Metrics are 850-hPa temperature (T), surface precipitation rate (SP) and precipitable water (PWAT).}
    \begin{tabular}{llcccc}\toprule
        & Run & \% Novelty & T ($\mathrm{K}$) & SP ($\mathrm{mm}/\mathrm{day}$) & PWAT ($\mathrm{kg}/\mathrm{m}^2$) \\ \midrule
        1 &  Baseline \eqref{eq:base} & 100\% & 2.09  & 1.78 & 2.79  \\ 
        2 & ML-corrected \eqref{eq:prog} with $g_{\mathrm{Tq}}$  & 0\% & 1.86 & 1.43  & 3.31 \\ 
        3 & ML-corrected with $g_{\mathrm{Tquv}}$ ($\star$) & 0\% &  2.43 & 3.39  & 5.33\\ 
        4 & ND ML \eqref{eq:nd} with $g_{\mathrm{Tq}}, \eta_{\mathrm{T, OCSVM}}$    & 2.5\% & 1.97  & 1.49& 3.65 \\ 
        5 & ND ML with $g_{\mathrm{Tquv}}, \eta_{\mathrm{T, minmax}}$  & 35.7\% & 5.15 & 3.57  & 10.14\\ 
        6 & ND ML with $g_{\mathrm{Tquv}}, \eta_{\mathrm{T, OCSVM}}$   & 40.0\% & 1.58  & 1.40 & 2.66 \\ 
        7 & ND ML with $g_{\mathrm{Tquv}}, \eta_{\mathrm{Tq, OCSVM}}$ & 50.7\% & \textbf{1.53} & \textbf{1.24} & \textbf{2.37} \\ \bottomrule
    \end{tabular}
    \label{tab:metrics}
\end{table}

% \begin{figure}
%     \centering
%     \includegraphics[width=0.8\textwidth]{fig/prevent-explosion.png}
%     \caption{RMSE of temperatures of 850hPa of pressure. 
%     While the novelty detection run with $\eta_{\mathrm{T, minmax}}$ is far worse than other approaches, it avoids the early crash of the ML-corrected run $g_{\mathrm{Tquv}}$.}
%     \label{fig:prevent-explosion}
% \end{figure}

% \begin{enumerate}
%     \item The \textbf{baseline run} evolves according to equation~\eqref{eq:base} with no ML corrections.
%     \item The \textbf{prognostic run without wind} follows equation~\eqref{eq:prog} with $g_{\mathrm{Tq}}$.
%     \item The \textbf{prognostic run with wind} follows equation~\eqref{eq:prog} with $g_{\mathrm{Tquv}}$.
%     \item The \textbf{prognostic run without wind and with OCSVM novelty detection} follows equation~\eqref{eq:nd} with $g_{\mathrm{Tq}}$ and $\eta_{\mathrm{OCSVM}}$ for RBF parameter $\gamma = \frac4{d}$ and cutoff $\nu = \smash{\min_{i', t'} \sum_{t} \sum_i \alpha_{t,i} \kappa_\gamma(x_{i'}^{(t')}, x_{i}^{(t)})}$ set to be the smallest OCSVM score observed in the training set.
%     \item The \textbf{prognostic run with wind and min-max novelty detection} follows equation~\eqref{eq:nd} with $g_{\mathrm{Tquv}}$ and $\eta_{\mathrm{minmax}}$.
%     \item The \textbf{prognostic run with wind and OCSVM novelty detection} follows equation~\eqref{eq:nd} with $g_{\mathrm{Tquv}}$ and $\eta_{\mathrm{OCSVM}}$.
% \end{enumerate}

\begin{figure}
    \centering
    \includegraphics[width=0.98\textwidth]{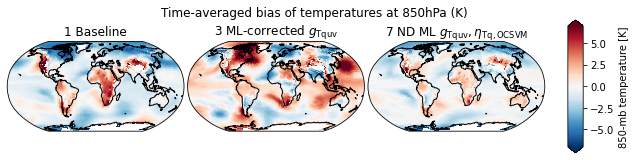}
    \caption{The OCSVM model (7, right) moderates the extreme behavior w.r.t. near-surface temperatures of both the baseline model (1, left) and the crashed ML-corrected run (3, center).}
    \label{fig:temp-map-plots}
\end{figure}

The baseline model (1) outperforms the ML-corrected model with winds without novelty detection (3).
In particular, the simulation of (3) crashed after 38 days due to model instability.
Applying novelty detection in (4) to (2) preserves the stability of the model without across-the-board improvements.
The min-max novelty detector (5) avoids the crash of (3), but otherwise performs far worse than other approaches, indicating the importance of representing the data distribution with a more meaningful quantity than a bounding box.
Both OCSVM novelty detection models with $g_{\mathrm{Tquv}}$ (6, 7) dominate all other models on all metrics, and the final model with $\eta_{\mathrm{Tq}}$ (7) performs best on all metrics.
These results suggest that suppressing ML corrections to atypical temperature or specific humidity columns is sufficient to realize the advantages of incorporating horizontal winds into the ML-corrected model, which ML-only models fail to achieve.
% These results suggest that the advantages to be attained by incorporating horizontal winds into the ML-corrected model can only be realized if the corrections are not applied to atypical temperature or specific humidity columns.
These models find a ``sweet spot'' between the ML-corrected and baseline approaches and stabilizes the ML corrections, reducing model bias without suffering from catastrophic behavior caused by out-of-sample errors, as illustrated for temperature in Figure~\ref{fig:temp-map-plots}.

\begin{figure}
    \centering
    \includegraphics[width=0.8\textwidth]{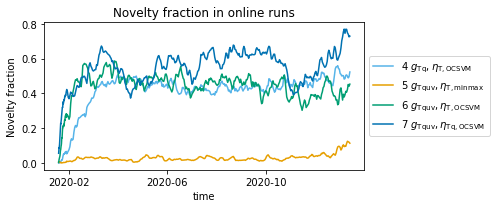}
    \caption{Percentage of novelties weighted by grid cell size for year-long runs (4-7).}
    \label{fig:novelty-frac}
\end{figure}

As visualized by Figure~\ref{fig:novelty-frac}, the number of novelties in each run with wind included in the ML corrections quickly spikes to include 40 to 70\% of the grid points (weighted by corresponding land area). 
The most successful approaches are thus highly aggressive, identifying a wide range of columns as novelties and removing them from consideration to erratic behavior.
While the novelty detection models flag a large fraction of columns, this behavior remains relatively stable over time, which indicates that these climate models are unlikely to converge to an ``effectively baseline'' solution, in which all columns are classified as novelties and ML corrections are never incorporated.

% improve upon both prognostic models and (6) obtains the best of each metric out of all models with much smaller biases than (1).
% The min-max novelty detector (4) avoids the crash of (3), but otherwise performs far worse than the others, which indicates that the rectangular region of support estimated by the latter deems destructive novelties as ``in-sample''; the OCSVM with RBF kernel allows a more compact estimation that categorizes more samples as novelties.

% The final two models suggest that novelty detection finds a ``sweet spot'' between the prognostic and baseline approaches and stabilizes the ML corrections, reducing model bias without suffering from catastrophic behavior caused by out-of-sample errors.
% Figure~\ref{fig:temp-map-plots} illustrates this in more detail by contrasting the time-averaged temperature errors of the (1), (6), and (3), showing how the OCSVM model tempers the extremal behavior of the baseline run over Canada and Africa and of the prognostic run over North America and the Antarctic.
We walk through an explicit example of how these novelty detectors preempt model instability in Appendix~\ref{asec:oos-catastrophe}.
% Appendix~\ref{asec:stability} conducts further stability analysis, by training several models with different random seeds and comparing the variances of each approach. 
Appendix~\ref{asec:ocsvm-param} shows that varying $\nu$ interpolates between the always-novelty (baseline) and never-novelty (ML-corrected) regimes for a variety of $\gamma$ choices and has optimal model performance between those extremes.

\section{Conclusion and future work}

Applying novelty detection to coarse ML-corrected atmospheric climate models improves the quality of temperature and precipitation estimates of coarse climate models by permitting the introduction of wind tendencies to the corrective model without instabilities.
Future work can build on this by experimenting with different novelty detection approaches, OCSVM kernels, inputs to $g$ and $\eta$, and methods of integration with the ML-corrected climate model.

% That said, this method is new and introduces a large space of hyperparameters to search: different methods (e.g. LOF, Gaussian) different kernels (e.g. polynomial), different restrictions (e.g. more or fewer SVs).
% This paper suggests that it's a fruitful space to explore, while leaving much of the exploration for future work.

\begin{ack}
This work was performed while the lead author was a summer intern at the Allen Institute for Artificial Intelligence (AI2).
We thank AI2 for supporting this work and NOAA-GFDL for running the 1-year X-SHiELD simulation on which our ML is trained using the Gaea computing system.
We also acknowledge NOAA-GFDL, NOAA-EMC, and the UFS community for making code and software packages publicly available.
We thank Daniel Hsu for helpful conversations.
\end{ack}

\bibliography{main}
\bibliographystyle{unsrt}

\appendix

\section{Training details for ML-correction $g$}\label{asec:training}

\subsection{Dataset}
We train the corrective tendencies offline as described by \cite{wbc21, bhk22}.
That is, $g(\cdot; \theta)$ is trained on samples \smash{$((x^{(t)}_i, q_i), y_i^{(t)})$} to ensure \smash{$g(x^{(t)}_i, q_i; \theta) \approx y^{(t)}_i$} for nudging tendency labels $y^{(t)}_i$ to be defined.
We obtain the samples by combining the results of two simulations of baseline climate models with $T$ time steps and $N$ grid cells:
\begin{itemize}
    \item $X^{(1)}, \dots, X^{(T)} \in \mathbb{R}^{N \times d}$ are a sequence of observed time steps of the nudged coarse run with $N$ grid cells described in \cite{wbc21} that that are corrected at each step with the observed nudging tencies $Y^{(1)}, \dots, Y^{(T)}$. 
    We use a version of FV3GFS \cite{zlcmcr19} with a C48 cubed-sphere grid of approximately 200~km horizontal resolution \cite{pl07} for our coarse-grid model $f_i$.
    In this grid, the Earth is divided into 6 square tiles with a 48-by-48 grid imposed on each ($N = 6 \cdot 48^2$). 
    \item $X_{\mathrm{fine}}^{(1)}, \dots, X_{\mathrm{fine}}^{(T)} \in \mathbb{R}^{N \times d}$ are generated by a simulating a fine-grid model with $N_{\mathrm{fine}} \gg N$ grid cells with equation~\eqref{eq:base} and coarsening its output to have resolution $N$. 
    The fine-grid reference model used by \cite{chm22} to train the ML has a similar type of grid with a much finer resolution of 3~km.
\end{itemize}
We scale the difference between the coarse model and its highly accurate fine-grained counterpart in order to obtain $d$-dimensional \textit{nudging tendencies}, \[y^{(t)}_i := \frac{x^{(t)}_{\mathrm{fine},i} - x_{i}^{(t)}}\tau,\] for nudging timescale $\tau$.
%%$f_i^{\mathrm{fine}}$ employs the C3072 (approx 3km) standard with 3072-by-3072 grids instead (used as the fine-resolution run in , with $N_{\mathrm{fine}} = 6 \cdot 3072^2$).
%Samples are collected by running a year-long simulation with $T = 365 \cdot 4$ time-steps at six-hour increments on both coarse and fine models, resulting in a total of $n = NT = 20183040$ training samples.
Samples are collected by running a year-long coarse-grid simulation nudged to the fine-grid model state with a 3 hour nudging timescale; the state and nudging tendencies are saved every 3 hours. After dividing this data into interleaved time blocks for the train/test split and subsampling down to 15\% of the columns in each timestep, we are left with $n = 2834611$ training samples.

The same dataset \smash{$\{x^{(t)}_i \in \mathbb{R}^d: t \in [T], i \in [N]\}$} is used to train the novelty detector $\eta(\cdot; \rho)$.
The nudges \smash{$y_i^{(t)}$} are omitted, as the novelty detection procedure requires no labels.

\subsection{ML-corrected climate models}
We consider two nudging tendency models: $g_{\mathrm{Tq}}$ and $g_{\mathrm{Tquv}}$.
\begin{itemize}
    \item $g_{\mathrm{Tq}}$ is a no-wind model that corrects vertical columns of air temperature (T) and specific humidity (q).
That is, $x_i$ is a $d = (2 \cdot 79)$-dimensional vector with 79 temperature and 79 humidity coordinates, each corresponding to a pressure-level in the atmosphere.
    \item $g_{\mathrm{Tquv}}$ also corrects horizontal wind tendencies (u, v), which modify the two-dimensional wind velocity at each altitude, making $x_i$ a $d = (4 \cdot 79)$-dimensional vector.
\end{itemize}

\smash{$g_{\mathrm{Tq}}(\cdot; \theta): \mathbb{R}^{158} \times \mathbb{R}^{3} \to \mathbb{R}^{158}$} predicts the temperature and humidity nudge vector \smash{$y_i^{(t)}$} from temperature and humidity state \smash{$x_i^{(t)}$}, as well as the insolation, surface elevation, and latitude $q_i$ of the corresponding cell.
We represent $g_{Tq}(\cdot; \theta)$ as a three-layer dense multi-layer perceptron of width 419.
The loss is measured by the mean absolute error (MAE) with kernel regularization with parameter $10^{-4}$.
We train the model with Adam for 500 epochs using a fixed learning rate of $0.00014$ and a batch size of 512 samples.
For the sake of stability, the model sets to zero the learned nudges for the three highest altitude temperature and humidity values; that is, $g_{\mathrm{Tq}}(\cdot; \theta)$ can be properly thought of as a function of the form \smash{$\mathbb{R}^{158} \times \mathbb{R}^{3} \to \mathbb{R}^{152}$}.

On the other hand, \smash{$g_{\mathrm{Tquv}}(\cdot; \theta): \mathbb{R}^{316} \times \mathbb{R}^{3} \to \mathbb{R}^{316}$} is defined as the concatenation of two learned functions for input $x = (x_{\mathrm{Tq}}, x_{\mathrm{uv}}) \in \mathbb{R}^{158} \times \mathbb{R}^{158}$:
\[g_{\mathrm{Tquv}}(x, q; \theta) = (g_{\mathrm{Tq}}(x_{\mathrm{Tq}}, q; \theta_{\mathrm{Tq}}), g_{\mathrm{uw}}(x_{\mathrm{Tq}},x_{\mathrm{uw}}, q; \theta_{\mathrm{uv}})).\]
$g_{\mathrm{Tq}}( \cdot; \theta_{\mathrm{Tq}})$ is trained identically to the aforementioned model. 
$g_{\mathrm{uv}}( \cdot; \theta_{\mathrm{uv}}): \mathbb{R}^{316} \times \mathbb{R}^{3} \to \mathbb{R}^{158}$ is separately trained to infer wind nudges from temperatures, humidities, and horizontal winds.
Besides the different input dimension, $g_{\mathrm{uv}}( \cdot; \theta_{\mathrm{uv}})$ is otherwise structured and trained identically to the other model.

\subsection{Computing scalar metrics}
As mentioned in Section~\ref{sec:results}, we measure the success of a coarse-grid simulated run by computing the RMSE of time-averaged quantities (850hPa temperature, surface precipitation, total precipitable water) with respect to those same quantities of the fine grid run.
We compute each with the following expression:
\begin{equation}\label{eq:rmse}
    \sqrt{ \sum_{i=1}^N a_i \left(\frac1T \sum_{t=1}^T \left(\hat{s}_{ i}^{(t)} - s_{\mathrm{fine},i}^{(t)}\right)\right)^2},
\end{equation}
letting $\hat{s}_i^{(t)}$ and $s_{\mathrm{fine}, i}^{(t)}$ reflect the quantity at grid cell $i \in [N]$ and time $t \in [T]$ in our coarse-grid and the reference fine-grid simulations respectively and $a_i$ represent the normalized area weights of grid cells.

\section{Comparing continuous and threshold-based novelty detectors $\eta(\cdot; \rho)$}\label{asec:threshold}

\begin{figure}
    \centering
    \includegraphics[width=0.9\textwidth]{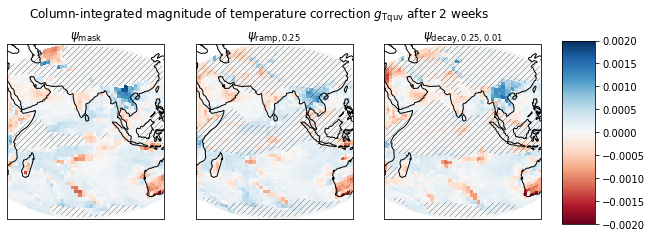}
    \caption{Integrated horizontal wind tendencies \textit{after novelty detection is applied} at a fixed time-step $t$ after two weeks of simulation on a simulation of equation~\eqref{eq:nd} with different choices of link  function $\psi_{\mathrm{mask}}$ (left), $\psi_{\mathrm{ramp}, 0.25}$ (center), and  $\psi_{\mathrm{decay}, -0.25, 0.01}$ (right) with $g_{\mathrm{Tquv}}$ and $\eta_{\mathrm{T, OCSVM}}$ with $\gamma = \frac4d$ trained on temperature tendencies.
    The hatched regions have some amount of ML-correction suppression (i.e. $\eta_{\mathrm{T, OCSVM}}(x; \rho) < 1$).
    % We use a solid line for the masking link function because it functions as a hard threshold, while the others have a dashed line to indicate the separation between no suppression and some amount of suppression.
    }
    \label{fig:smooth-nd-map}
\end{figure}

The results in Section~\ref{sec:results} demonstrate the empirical success of the OCSVM novelty detector applied as a discrete on-off switch for the corrective tendencies in equation~\eqref{eq:nd}.
However, this method creates sharp discontinuities in the nudging tendencies (see the left panel of Figure~\ref{fig:smooth-nd-map}), since the tendencies are often at their most extreme when the respective temperature columns are nearly out of sample. 
As a potential remedy, we consider several approaches to smoothing these sharp novelty detection functions $\eta$ and ask whether they (1) subjectively smooth the tendencies to avoid such sharp thresholds and (2) result in better (at least, not worse) online model performance.

We represent the generalized OCSVM novelty detector applied to temperature columns as 
\[\eta_{\mathrm{T, OCSVM}}(x; (\alpha, \nu, \gamma, \psi)) = \psi\left(\sum_{t=1}^T \sum_{i=1}^N \alpha_{t,i} \kappa_\gamma(x, x_{i}^{(t)}) - \nu\right),\]
for some \textit{link function} $\psi: \mathbb{R} \to [0, 1]$.
We consider three choices for $\psi$.
\begin{enumerate}
    \item $\psi_{\mathrm{mask}}$ is the sharp threshold function used in the paper body: 
    \[\psi_{\mathrm{mask}}(z) = \begin{cases}
        1 &  \text{if } z \geq 0, \\
        0 & \text{if } z < 0.
    \end{cases} \]
    \item $\psi_{\mathrm{ramp}, \zeta}$ linearly interpolates between 1 at $\zeta > 0$ and 0 at $-\zeta$:
    \[\psi_{\mathrm{ramp}, \zeta}(z) = \begin{cases}
        1 & \text{if } z \geq \zeta, \\
        \frac{z + \zeta}{2 \zeta}\ & \text{if } z \in [-\zeta, \zeta],\\
        0 & \text{if } z \leq -\zeta.
    \end{cases}
    \]
    \item $\psi_{\mathrm{decay}, \zeta, \beta}$ exponentially decays at a rate of $\beta \in (0, 1)$ starting at the threshold $\zeta$:
    \[\psi_{\mathrm{decay}, \zeta, \beta}(z) = \begin{cases}
        1 & \text{if } z \geq \zeta, \\
        \beta^{\zeta - z} & \text{if } z \leq \zeta.
    \end{cases}\]
\end{enumerate}
While numerous other sigmoidal functions can be considered, we restrict our focus to these three.
We use the same ML correction $g_{\mathrm{Tquv}}$ and OCSVM model with trained $\alpha$ and $\gamma = \frac{4}{79}$ and compare the performances of several simulated runs of equation~\eqref{eq:nd} with different link functions $\psi$.

Table~\ref{tab:metrics-thres} illustrates that the choice of link function $\psi$ has a marginal impact on the scalar quality metrics, with the ramp function besting the masked approach at temperature and precipitation accuracy while falling short on precipitable water.
Figure~\ref{fig:smooth-nd-map} shows that these link functions have a locally smoothing effect on the magnitudes of the learned tendencies; both ramp and decay link functions avoid the sharp separations between complete suppression and expression of strong corrections from the masking link function. 

% \red{Figure~\ref{fig:smooth-nd}: Mask vs ramp vs decay ND metrics for OCSVM with 4/79}

\begin{table}[]
    \centering
    \caption{A comparison of RMSEs of time-averaged metrics over year-long simulations of equation \eqref{eq:nd} with different choices of link function $\psi$ with $g_{\mathrm{Tquv}}$ and $\eta_{\mathrm{T, ocsvm}}$ as discussed in Section~\ref{sec:results}.}
    \begin{tabular}{llccc}\toprule
        & Run & T ($\mathrm{K}$) & SP ($\mathrm{mm}/\mathrm{day}$) & PWAT ($\mathrm{kg}/\mathrm{m}^2$) \\ \midrule
        1& $\psi_{\mathrm{mask}}$ & 1.58 & 1.40  & \textbf{2.66} \\ 
        2&$\psi_{\mathrm{ramp}, 0.1}$ & 1.49  & {1.36}   & 2.73 \\ 
        3&$\psi_{\mathrm{ramp}, 0.25}$ & 1.53  & {1.36}   & 2.71 \\ 
        4&$\psi_{\mathrm{ramp}, 0.5}$ & \textbf{1.45}  & 1.39   & 2.69  \\ 
        5&$\psi_{\mathrm{decay}, 0.25, 0.005}$ & 1.48  & \textbf{1.34}  & 2.69  \\ 
        6&$\psi_{\mathrm{decay}, 0.25, 0.01}$ & 1.53 & 1.42   & 2.83  \\ 
        7&$\psi_{\mathrm{decay}, 0.25, 0.02}$ & 1.61  & 1.43   & 2.82  \\ \bottomrule
    \end{tabular}
    \label{tab:metrics-thres}
\end{table}

% \begin{figure}
%     \centering
%     \includegraphics[width=0.9\textwidth]{fig/smooth-nd-temp.png}
%     \caption{\red{todo}}
%     \label{fig:smooth-nd}
% \end{figure}

\section{An instance of novelty detectors preventing a catastrophic out-of-sample error}\label{asec:oos-catastrophe}

\begin{figure}
    \centering
    \includegraphics[width=\textwidth]{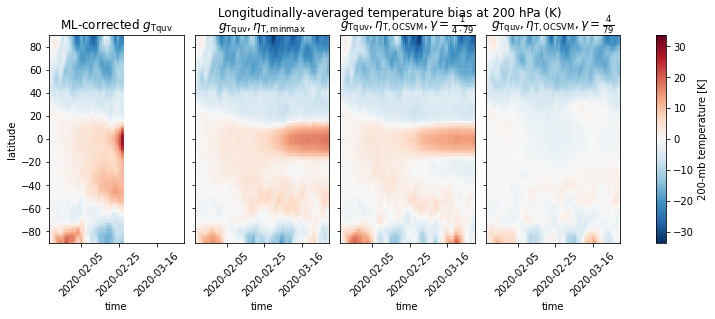}
    \includegraphics[width=\textwidth]{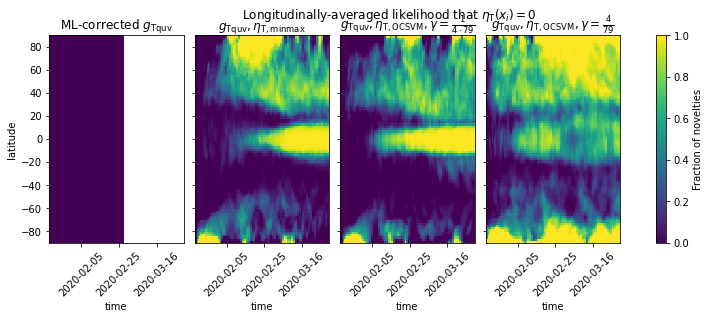}
    \caption{Hovmoller (time vs latitude) plots visualizing projected upper-atmospheric temperature biases (against a fine-grid reference simulation) at the 200hPa pressure level (top) and fractions of novelties identified (bottom) by four different models (left to right): (1) the ML-corrected climate model $g_{\mathrm{Tquv}}$ without novelty detection (which crashes after 38 days), (2) $g_{\mathrm{Tquv}}$ with min-max novelty detection $\eta_{\mathrm{T, minmax}}$, (3) $g_{\mathrm{Tquv}}$ with OCSVM temperature novelty detection $\eta_{\mathrm{T, OCSVM}}$ with small \smash{$\gamma = \frac1{4 \cdot 79}$}, and (4) $g_{\mathrm{Tquv}}$ with OCSVM temperature novelty detection $\eta_{\mathrm{T, OCSVM}}$ with large \smash{$\gamma = \frac4{ 79}$}.
    (1) crashes due to an increase in upper atmospheric tropic temperatures, while (2) and (3) prevent equatorial temperature increases from spiraling out of control by consistently identifying novelties in that region. 
    (4) identifies these irregularities early enough to prevent the temperature shift outright.}
    \label{fig:temp-explosion}
\end{figure}

This appendix walks through a specific example of how novelty detection approach prevents a model from crashing.
Figure~\ref{fig:temp-explosion} compares the first ten weeks of four ML-corrected simulations: one without novelty detection (equation~\eqref{eq:prog}) and three with novelty detectors (equation~\eqref{eq:nd}; a min-max novelty detector and two OCSVMs with different choices of smoothness parameter $\gamma$).

The ML-corrected model without novelty detection crashes after 38 days due to an explosion of equatorial upper-atmospheric temperatures, as viewed in the top-left plot.
While the spike in temperatures occurs right before the crash, the model predicted that tropical regions would be hotter than expected before that point.
Since tropical regions are among the hottest, this initial heat shift indicates that the predicted temperature columns are likely to be hotter than anything observed in the training dataset.

This narrative is supported by the fact that the models with min-max and OCSVM with $\gamma = \frac1{316}$ novelty detectors do not crash.
Both are faced with heated equatorial columns at the beginning of the simulation, and those are identified as out-of-sample by both novelty detectors before the aforementioned temperature spike (see the bottom plots).
By removing the ML-corrected nudge from the equatorial region, the equatorial temperature biases are bounded and persist due to nearly every equatorial column being identified as a novelty.

On the other hand, the OCSVM with $\gamma = \frac4{79}$ entirely negates the equatorial temperature bias by identifying the shift as a novelty far early in time than any other model.
While this approach occasionally identifies some tropical columns as out-of-sample during the run, it can reap the benefits of ML-nudging much of the time near the equator.

% \section{Model stability comparison}\label{asec:stability}

% \red{Run several runs with each seed and compute variances/ranges of various metrics}

\section{OCSVM parameter comparison}\label{asec:ocsvm-param}

\begin{figure}
    \centering
    \includegraphics[width=0.45\textwidth]{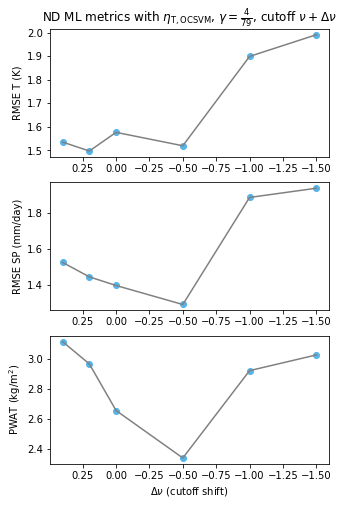}
    \includegraphics[width=0.45\textwidth]{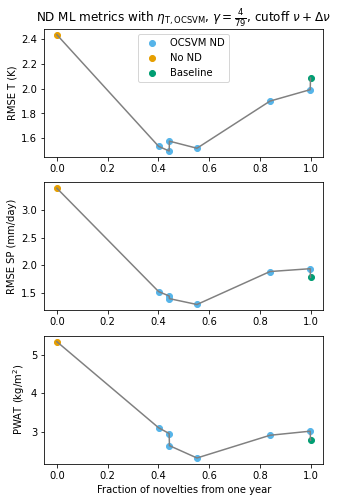}
    \caption{
    Compares the RMSE of time-averaged 850hPa temperature (top), surface precipitation (center), and precipitable water (bottom) of several year-long simulations of equation~\eqref{eq:nd} with ML-correction $g_{\mathrm{Tquv}}$ and novelty detector $\eta_{\mathrm{T, OCSVM}}$ with $\gamma = \frac4{79}$ and cutoff $\nu + \Delta \nu$ where $\nu$ is obtained by equation~\eqref{eq:nu} for $\Delta\nu \in \{-1.5, -1, -0.5, 0, 0.2, 0.4\}$.
    The left plots show the metric as a function of the cutoff.
    The right plots instead have the total fraction of novelties identified (which is monotonic as a function of the cutoff) on the x-axis.
    The right plots additionally feature the the crashed result of the ML-corrected run equation~\eqref{eq:prog} with $g_{\mathrm{Tquv}}$ without novelty detection as a         ``never-novelty'' run and the baseline result of equation~\eqref{eq:base} as an ``always-novelty'' run.
    }
    \label{fig:params}
\end{figure}

Section~\ref{sec:results} considers an OCSVM with $\gamma = \frac4{d}$ and $\nu$ set to the minimum observed score in the training data and argues that the this model applied to either only temperature or both temperature and humid finds the ``sweet spot'' between the baseline run and the full ML-corrected run.
Here, we validate that conclusion by considering several choices of $\gamma$ and varying $\nu$ to adjust the sensitivity of the novelty detector.
We show that these approaches interpolate between the baseline and ML-corrected run as the cutoffs change and that the metrics are optimized by choosing an intermediate model that categorizes a substantial fraction of samples as novelties.

In Figure~\ref{fig:params}, we consider an ML corrected model $g_{\mathrm{Tquv}}$ augmented with an OCSVM novelty detector $\eta_{\mathrm{T, OCSVM}}$ with $\gamma = \frac4{79}$ and various choices of cutoff.
When the scalar metrics for evaluating a year-long run are plotted as a function of the cutoff, we find that an intermediate cutoff choice yields optimal performance.
We similarly plot these metrics as a function of the fraction of novelties identified by each cutoff and observe a curve with a local minimum that occurs when between 40\% and 60\% of all samples are deemed novelties. 
This plot---which includes visualizations of the skill of the crashed ML-corrected run (without novelty detection) and the baseline run---demonstrates that this approach effectively interpolates between those two extreme methods and that the cutoff $\nu$ from equation~\eqref{eq:nu} lies near that sweet spot.

\begin{figure}
    \centering
    \includegraphics[width=0.45\textwidth]{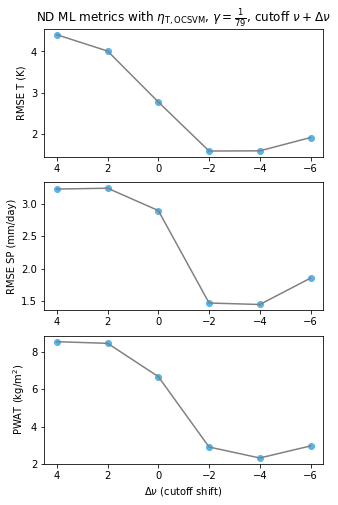}
    \includegraphics[width=0.45\textwidth]{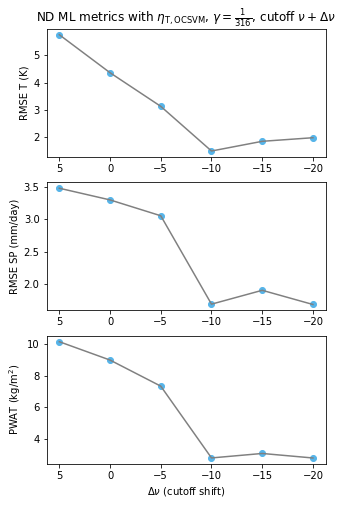}
    \caption{
    Replicates the experiment visualized in the left panel of Figure~\ref{fig:params}, but instead considers OCSVM novelty detectors with $\gamma = \frac{1}{79}$ and $\gamma = \frac1{316}$ and $\delta\nu \in \{-6, -4, -2, 0, 2, 4\}$ and $\delta\nu \in \{-20, -15, -10, -5, 0, 5\}$ respectively.
    }
    \label{fig:ocsvm-param}
\end{figure}

Moreover, this behavior is not isolated to the specific OCSVM considered here and elsewhere in the paper.
We train two additional OCSVM models with $\gamma \in \{\frac1{79}, \frac1{316}\}$ and similarly consider a wide range of cutoffs, which are plotted in Figure~\eqref{fig:ocsvm-param}.
We find that intermediate choices of the cutoff (at roughly $-4$ and $-10$ respectively) lead to better model RMSE scores on the time-averaged scalar metrics.

\end{document}